\documentstyle[preprint,aps,multicol,psfig,colordvi]{revtex} 
\def\beq{\begin{equation}} 
\def\eeq{\end{equation}} 
\def\beqa{\begin{eqnarray}} 
\def\eeqa{\end{eqnarray}} 
\def\l{\left} 
\def\r{\right} 
\def\bdi{\begin{displaymath}} 
\def\edi{\end{displaymath}} 
 
\begin{document} 
\tightenlines 
\title{DNA sequence from the unzipping force? : one mutation problem} 

\author{Somendra M. Bhattacharjee\dag\cite{eml1}} 

\address{\dag Institute of Physics, Bhubaneswar 751 005, India}  

\author{D. Marenduzzo\ddag\cite{eml1}} 

\address{\ddag International School for Advanced Studies (SISSA) 
and INFM, Via Beirut 2-4, 34014, Trieste, Italy}  


\maketitle

 
\begin{abstract} 
 The possibility of detecting mutations in a DNA from force measurements 
 (as a first step towards sequence analysis) is 
 discussed theoretically based on exact calculations.
 The force signal is associated with the domain wall separating  
 the zipped from the unzipped regions. We propose a comparison
 method (``differential force microscope'') to detect mutations.
 Two lattice models are treated as specific examples.
\end{abstract}

\noindent{87.14Gg, 87.80Fe, 05.20Gg} 
 

\newpage
 
The possibility of a force induced unzipping transition
\cite{Boc1,Boc2,smb,Nel,maren,gerland} has opened up new ways of
exploring the properties of biomolecules.  Since the critical force
for a double stranded DNA depends on its sequence, an inverse
problem can be posed: ``Can the DNA sequence be detected from the
force required to unzip it ?'' A still simpler question would be:
``Can the mutations in a DNA be detected by force measurements?''. A
positive answer to either of these questions would lead to the
possibility of testing (and detecting) mutations and of sequence
determination in a non-destructive way.
 
Based on a few simple models used earlier for the DNA denaturation and
unzipping transitions\cite{smb,Nel,maren}, we show here the signature
of mutations on the $f$-vs-$r$ curve (force versus relative distance
of the end points of the two strands).  The inverse problem is then to
get the position of the mutations from such an experimentally
realizable curve.  Our emphasis in this paper is on the one base pair
mutation problem also known as point mutation.  This is not just of
academic interest. The replication of DNA is a high-fidelity process,
thanks to the inbuilt proof-reading and repair mechanisms, so that
mistakes are very rare though even one could play havoc.

Our proposal (solution of the inverse problem), based on exact
calculations, is to obtain the force difference (``differential
force'') between identically stretched native and mutant DNAs.  This
could be done by using {\em e.g.} two atomic force microscope tips: we
name this apparatus a ``differential force microscope''. The position
of the mutation can be obtained from a calibration curve involving the
extremum position (or its value) of the differential force curve. In
our models, we can find the nature of the mutation from the sign of
the differential force.
 
Let us model a double stranded (ds) DNA by two N-monomer polymers
interacting at the same contour length (or monomer index) $j$ of the
strands through a contact attractive potential 
$-\epsilon_j(\epsilon_j>0)$, which might depend on the contour length.  The
interaction energy is 
$H=-\sum_{j=1}^N \delta_{{\bf r}_j,{\bf    0}}\epsilon_j$, 
where ${\bf r}_j$ denotes the relative distance of
the $j-$th base pair and $\delta$ denotes the Kronecker delta.  Other
features like the self- and mutual-avoidance, base stacking energy,
helicity, etc are ignored in this study (but see below) in order to
focus on the base pairing energy. Such a model exhibits a denaturation
transition at a model-dependent temperature $T=T_m$\cite{den} from a
low $T$ double stranded configuration to a high $T$ phase of two
unpaired single strands.
 
A few definitions: Starting with a sequence
$\{\epsilon_j|j=1,\ldots,N\}$ or bases of a DNA (to be called the
native DNA), we define a mutant DNA as one with almost the same base
sequence as the native molecule {\it except } a few base pairs.  A
homogeneous DNA or homo-DNA is a DNA with identical base pairs (i.e.
$\epsilon_j =\epsilon$ for all $j$) while a heterogeneous DNA is one
with heterogeneity in the sequence ($j$-dependent $\epsilon_j$).
Notice that the model we have defined does not consider base pair
stacking interaction and therefore does not distinguish {\em e.g.}
between an AT and a TA (or CG and GC) base pair.  Consequently, the
mutations we are referring to are only those involving AT (or TA)
$\leftrightarrow $ CG (or GC).
 
Two specific examples are considered here because of their exact
tractability (analytical and numerical): (Mi) a two dimensional $d=2$
directed DNA model (with the strands directed along $(1,1)$) with
base-pair interaction and mutual avoidance (forbidden crossing of the
strands), and (Mii) Two Gaussian polymers with the base-pair
interaction in $d$-dimensions.  In both cases, only the relative
chain, involving the separation {\bf r} of the bases at the same
contour length, need be considered.  These models in addition to the
denaturation transition also show, for $T<T_m$, an unzipping
transition in presence of a force at the free end
($j=N$)\cite{smb,Nel,maren}.  In this paper, we consider the conjugate
fixed distance ensemble and henceforth restrict ourselves to $T<T_m$.
Consequently, we put $\beta=(k_B T)^{-1}=1$ unless explicitly shown,
where $k_B$ is the Boltzmann constant.

\renewcommand{\thefootnote}{\fnsymbol{footnote}}
 
For any ds DNA having their first monomers ($j=1$) joined and their
last monomers ($j=N$) at a relative lattice distance ${\bf r}$, the
force ${\bf f}_N({\bf r})$ required to maintain this relative distance
${\bf r}$ is ${\bf f}_N({\bf r})\equiv \nabla {\cal F}({\bf r})$ where
${\cal F} ({\bf r})$ represents the free energy of the system in the
fixed-${\bf r}$ ensemble$^{\footnotemark[4]}$.
\footnotetext[4]{In the continuum the force for the Gaussian 
interacting polymers satisfies a Burgers' 
type equation.  See, e.g.,  Ref. \cite{note1}.}  
By definition, 
$\int_0^\infty{\bf f}_N({\bf r})\cdot d {\bf r}$ =
${\cal  F}(\infty)-{\cal F}(0)$  which gives the free energy of
binding or the work required to unzip completely DNA.
 
The two ensembles, fixed-force and fixed-stretch, are expected to give
identical results in the $N\rightarrow \infty$ limit, though for
finite $N$ inequivalence might be expected \cite{Neu} (this is indeed
the picture valid for the homo-DNA), but a more serious situation
arises if, {\it e.g.} for the scalar case, $\partial f/\partial x <0$
in the fixed-stretch ensemble because, in the fixed-force ensemble,
$\frac{\partial\langle x\rangle}{\partial \beta f}$ 
$\equiv \langle x^2\rangle-\langle x\rangle^2\ge 0$, where 
$\langle\cdot\rangle$
denotes thermal averaging.  This is the case for a heterogeneous DNA
as shown in Fig. 1.  The regions of ``wrong'' sign (reminiscent of
preshocks in Burgers turbulence\cite{note1} or of ``slip'' in
Ref.\cite{Boc2}) go away only after quenched averaging but do survive
in the thermodynamic limit for each individual realization.  {\it The
  absence of self-averaging in the system is encouraging, because it
  implies that individual features, typical of a single realization of
  the sequence, are maintained in the characteristic curves.}  We draw
attention to the overlap of the $f$ vs $x$ curves over a range of $x$
in Fig. 1 for different lengths with identical sequence near the open
end.  This overlap is a sign that the modulation is a characteristic
of the sequence, and we have found this to be true quite generally.
 
Let us  now consider the one mutation problem where the  
$k$-th pairing energy of the native DNA  has been changed from  
$\epsilon_k$ to $\epsilon'_k$.    
For this one site change in $H$, the 
partition function $Z_{N|k}({\bf r},N)$ is given by 
\beq\label{mutation_one} 
Z_{N|k}({\bf r},N)=Z_N({\bf r},N)+c_k Z_N({\bf r},N\mid {\bf 0},k). 
\eeq 
where $c_k\equiv\l(\exp{\l(\beta(\epsilon'_k-\epsilon_k)\r)}-1\r)$ and 
 $Z_N({\bf r},N)$  is the native DNA partition function   
with last monomers at a relative distance ${\bf r}$, while 
$Z_{N}({\bf r},N\mid {\bf 0},k)$ is with the 
additional constraint of the $k-$th pair being zipped.   
Notice that Eq. \ref{mutation_one} 
 applies for both heterogeneous and homogeneous cases 
(and also for self avoiding strands).  
 
We define a zipping probability $P({\bf r},N|{\bf 0},k)\equiv$
$\frac{Z_{N}({\bf r},N|{\bf 0},k)} {Z_N({\bf r},N)}$, which is the
conditional probability that site $k$ is zipped when the free ends are
at a distance ${\bf r}$.  The force difference (to be called the
differential force) to keep the free ends of both the mutant and the
native DNA at the same distance ${\bf r}$ can be written as 
\begin{eqnarray} 
\delta {\bf f}_{N|k}(\bf r)&=&-\ \frac{c_k}{1+c_kP({\bf r},N|{\bf 0},k)} 
 \nabla 
 P({\bf r},N|{\bf 0},k)\\ 
& =&  
\frac{c_kP({\bf r},N|{\bf 0},k)}{1+c_kP({\bf r},N|{\bf 0},k)} 
({\bf f}_{N}({\bf r}|{\bf 0},k)-{\bf f}_{N}({\bf r})). 
\end{eqnarray} 
\label{mut3} 

Here ${\bf f}_{N} ({\bf r}|{\bf 0},k)$ is a generalized force, the one
necessary to keep the free ends of the native DNA at a relative
distance ${\bf r}$ when the $k-$th monomers are zipped.  Except $c_k$,
all other quantities in Eq. \ref{mut3} refer to the native DNA.  This
fact allows the inverse problem to be tackled as we show explicitly
for a few cases.  Since for DNA we have generally two possible choices
for $\epsilon$, ($\epsilon_1$ and $\epsilon_2$), the sign of $c_k$
determines the sign of $\delta f$ in Eq. \ref{mut3}.
Thus, in our simple model,  the nature of the mutation can be identified 
from the sign of the differential force curve.
Eqs. \ref{mutation_one},\ref{mut3} can be generalized to more than one
mutations and to models with other local energy
parameters,
 though they become algebraically more
involved. These more complicated cases will be discussed elsewhere. We
consider the simplest case here.
 
If ${\bf r}\equiv (x,0,\ldots,0)$ is the direction of stretching, the
quantity $P({\bf r},N|{\bf 0},k)$ has a kink-like behavior (as in Fig.
\ref{fig:2}).  In the fixed stretch ensemble, the chain separates into
an unzipped and a zipped region separated by a domain wall, which to a
very good approximation can be fit by a $\tanh$ function:
  
\begin{eqnarray} 
  \label{eq:3} 
P({\bf  r},N&|&{\bf 0},k)\equiv P({\bf 0},N|{\bf 0},k)e^{-\beta 
\int d{\bf r}\cdot ({\bf  
   f}_{N}({\bf r}|{\bf 0},k)- {\bf f}_N({\bf r}))} \nonumber\\ 
&\approx&P_0\l(1+\tanh{\chi}\r)/2, \,  
\chi\equiv{\l[x-x_{\rm d}(k)\r]} 
/{w_{\rm d}(k)}, 
\end{eqnarray} 
where $x_{\rm d}(k)$ and $w_{\rm d}(k)$ are the position and the width
of the wall (kink), and $P_0$ is a constant.  Eq. \ref{eq:3} suggests
that $\delta f(x)\equiv w_{\rm d}(k)^{-1} {\tilde {\sf f}}(\chi)$,
where ${\tilde {\sf f}}(\chi)$ is a scaling function.
 
It is possible to extend the above analysis to the general case where
more than one mutation is present.  For example, the partition
function for the case with two mutations at positions $k_1$ and $k_2$,
in an obvious notation, is
\begin{eqnarray}
  \label{eq:4}
  Z_{N|k_1,k_2}({\bf r})=Z_N({\bf
  r})&+&\sum_{i=1,2}c_{k_i}Z_{N}({\bf r}N|{\bf 0}k_i)+\nonumber\\
&&c_{k_1}c_{k_2}Z_{N}({\bf r}N|{\bf 0}k_1|{\bf 0}k_2).
\end{eqnarray}
The last term of the above equation, representing correlation of the
mutations, gives an additional contribution to the differential force
over and above the individual contributions of the mutations.  This
additional contribution is negligible if the two mutations are far
away or, more quantitatively, not in the same domain wall. 

We now use these general results for cases (Mi),(Mii). 
 
In the two dimensional model (Mi), the partition function for 
two directed chains 
 having their last monomers at a relative lattice distance 
$x$ (along $(1,-1)$ and in unit of the elementary square diagonal), 
 and their first monomers joined, can be written in terms of N 
monomer-to-monomer transfer matrices $W_j$ ($j=1,\ldots,N$) 
with matrix elements  
$\langle x'|W_j|x\rangle\equiv$ 
$\l(\l(\exp{\l(\beta\epsilon_j\r)}-1\r)\delta_{x',0}+1\r)$ 
$(2\delta_{x',x}+\delta_{x',x+1}+\delta_{x',x-1})$  
$|x\rangle,|x'\rangle$ being the position vectors (with the constraint  
$x,x'\ge 0$). 
 
For a homo-DNA with contact energy $\epsilon$, the largest eigenvalue
of $W$ determines the free energy and the thermodynamic properties in
the limit $N\rightarrow\infty$. For 
$T<T_m\equiv\frac{\epsilon}{k_B\log{\frac{4}{3}}}$, the melting temperature,
$\beta f(x)=\cosh^{-1} (\frac{1}{2z_0}-1)$ with 
$z_0=\sqrt{X} -X,X\equiv(1-e^{-\beta\epsilon})$.  Indeed, in the fixed-stretch
ensemble, any finite ($x\ll N$) stretch puts the chain on the phase
coexistence curve.
 
Exploiting the equivalence of the ensembles valid for for homo-DNA ,
we find $\beta f_N(x)\equiv \hat{F}(x/N)$, where
\beqa
\label{characteristic} 
\hat{F}(y)&=&2 \tanh^{-1}(\max\{y,y_0\}),\quad ( y_0\equiv
{\l(1-4z_0\r)}^{\frac{1}{2}}), 
\eeqa 
is a piecewise continuous
non-analytic function. For finite $N$, there is no singularity but the
approximation of Eq. \ref{characteristic} still works quite well. At
$x\sim y_0 N$ the force curve increases sharply (see also Fig.1,
curves (a) and (c)).
 
We now come to the explicit results for the one-mutation case where 
one $\epsilon$ is replaced by  $\epsilon' <\epsilon$.  
For a homo-DNA, by starting from 
Eq. \ref{characteristic}, using Eqs. \ref{mut3},\ref{eq:3},
one can find  analytical approximations to 
the shapes of the previously introduced $P(x,N|0,k)$ and $\delta f$ in 
terms of piecewise continuous functions.  We find {\em e.g.} 
$P(x,N|0,k)$ = ${P(0,N|0,k)} \exp{\l(g(x))\r)}$, where (if $Ny_0<\bar{k} 
\equiv N-k$): 
\begin{eqnarray}   
g(x)&=&0 \quad {\rm when}\  x<\bar{k}y_0,\\  
&=&g_{\bar{k}}(x) \ {\rm if} \ \bar{k}y_0<x<Ny_0 \\ 
&=&g_{\bar{k}}(x)-g_N(x) \ {\rm if} \ Ny_0<x<\bar{k}   
\end{eqnarray} 
\label{eq:2} 
and $g_{{k}}(x)=\log{\l[\l(\frac{1+\frac{x}{{k}}}{1+y_0}\r)^ {x+{k}} 
 \l(\frac{1-y_0}{1-\frac{x} {{k}}}\r)^{x-{k}}\r]}$.  
Eq. \ref{mut3} now 
simplifies because $f_{N}(x|0,k)=f_{\bar{k}}(x)$.   
The characteristics of the differential force curve $\delta f$ vs $x$, such 
as the extremum value, its position and the width,  
$\delta f_{\rm max}(k)$,  
$x_f(k)$ and $w_f(k)$,  can be determined from Eqs. \ref{eq:2} as 
\beq\label{results}  
\delta f_{\rm max}(k){\sim} [w_f(k)]^{-1}  
\sim {\bar{k}}^{-1/2}, x_f(k) 
={\bar{k}}{\tilde x}_f(\bar{k})  
\eeq  
with ${\tilde x}_f(0)=1$ and 
$\lim_{\bar{k}\to\infty}{\tilde x}_f (\bar{k})=y_0$, 
where $y_0$ is defined in Eq. \ref{characteristic}.  The 
area of the peak, which yields the difference (with respect to the 
native case) in the work necessary to unzip 
 completely the molecule, is 
constant as expected.  The scaling form introduced after 
Eq. \ref{eq:3} suggests that the differential force is significant 
only in the domain wall region  and the width of the domain wall   
$w_{\rm d}(k) \sim w_f(k)$ as we do see in the numerical results. 
 
For model (Mii), with one-dimensional Gaussian polymers
($T_m=\infty$), $P(x,N|0,k)$ has been calculated exactly by a transfer
matrix method and is shown in Fig.2a.  The validity of Eq.
\ref{results} ($\delta f_{\rm max}(k)\sim w_f(k)^{-1}$) is apparent
from the data-collapse of the various $\delta f$ curves in Fig.2b
where $\frac{\delta f}{\delta f_{\rm max}(k)}$ is plotted against
$(x-x_f(k)) \delta f_{\rm max}(k)$.  The peak force difference,
$\delta f_{\rm max}(k)$ as a function of the mutant position $k$ is in
accord with the $\bar{k}^{-1/2}$ law of Eq. \ref{results}. The results
for model (Mi) are shown in Fig. \ref{fig:3}.  For $d>1$ (${\bf
  r}=(x,0,\ldots,0)$ as above) the situation is similar as, {\it
  e.g.}, the data-collapse of Fig. \ref{fig:2} and Eq. \ref{results}
remain valid.

Coming to the case of one mutation on a heterogeneous DNA consisting
of two energies $\epsilon_1$ and $\epsilon_2 >\epsilon_1$ chosen with
equal probability, the shapes of the zipping probability curves are
found to be similar to the homo-DNA case, in fact indistinguishable on
the plot of Fig. \ref{fig:2}, with $x_{\rm d}(k)$ and $w_{\rm d}(k)$
sequence dependent. This indicates the validity of the domain wall
interpretation even for a heterogeneous DNA.  The mutation involves a
change of the energy at site $k$ (i.e. $\epsilon_1 \leftrightarrow
\epsilon_2$).  The signals $\delta f$ for various mutations are shown
in Fig. \ref{fig:4}.  As already mentioned, in our models the sign of
$\delta f$ tells us the nature of the mutation. These individual
curves can again be collapsed on to a single one as for homo-DNA,
though the nature of collapse is not as good, mainly because the area
under the curve is no longer strictly a constant.  This reflects the
importance of local sequences around the mutation point.  Figure 3
(curves b, c) shows the $k$-dependence of $x_f(k)$.  Unlike for
homo-DNA, $\delta f_{\rm max}(k)$ does not seem to have the simple
form of curve (a) in Fig. 3.  Although the linearity is maintained for
$x_f(k)$ as for the homogeneous case, there are regions of
nonmonotonicity at small scales which hamper the inversion.

Fig. 3 gives a basis for a calibration curve. This could be $x_f(k)$
or $\delta f_{\rm max}$ vs k (or both) for a homogeneous DNA, though
for heterogeneous DNA, we find the $x_f$-vs-$k$ curve to be more
reliable. Given the value of $x_f(k)$, one can look up in Fig.3 for
the corresponding $k$.  The accuracy of the method relies on the
ability to resolve close-by mutation points, {\em i.e.} mutation
points in the same domain wall.  There are differences in the full
profiles of the $\delta f$ curves for mutations at, say, $k$, $k+1$,
$k+2$, though translating that information back to the identification
of the position is yet to be achieved. A better resolution
is in any case obtained by changing the point at which the strands are
pulled.

We now argue on how our calculation can compare with a potential
experiment.  In \cite{Boc1,Boc2}, the typical force arising in the
unzipping is between $10$ and $15$ pN, while the resolution is set
below $0.2$ pN, so in percentage it is $<1.3-2\%$.  Dynamical effects
(important in \cite{peyrard}) are almost negligible already at the
lowest stretching velocity used in \cite{Boc1,Boc2} ($20$ nm/s) and
are less important as the velocity is lowered.  Our values for the
typical force at $T=1$ are at the border of this present day
resolution (see Fig. 3, where it appears that the resolution is around
$1\%$ for $k\sim 500$, {\it i.e.} on the middle of the chains).
In principle they can be improved above it by lowering $T$,
though it is not clear to what extent this will work because
the experimental temperature range is rather limited. 
In Ref.\cite{thompson-siggia}, the authors suggest that there will be experimental 
difficulties which would hamper the acquisition of the base by base sequence of DNA by
means of force measurements (but would however allow to get
information over groups of $\sim 10$ bases). This difficulty, though absent in our 
exact analysis of the models,  might also set a  lower limit on the error on the 
position of the mutation. Summarizing, we prove that mutations are detectable in the
theoretical models.
Numbers coming from our models suggest that this measurement might be a benchmark
for present day real technology.

We finally propose an algorithm for
sequencing DNA from the unzipping force in our models.
This is defined so that the energy of
the ${j}-$th base pair is $\epsilon_2$ if the average
force at stretch $x=N-j$ is above
the force signal of a homo-DNA with an attractive energy
$\epsilon\equiv\frac{\epsilon_1+\epsilon_2}{2}$
($T$ is low enough that $y_0\sim 1$).
Our algorithm differs from the discussion in
\cite{Boc2} because $T$ and extra constraints (see below)
play crucial roles.
If we define the ``score''
of the algorithm as the fraction of base pairs correctly
predicted, from our data
we observe that for any finite-size sample  the score is 100\%
for $T<T_0(N){\sim}{N}^{-\psi}$, with $0\le\psi\le 1$
for $N\to\infty$.
However, $N_0$ monomers near the open end can be
sequenced at $T\simeq T_0(N_0)$, no matter how big the total $N$ is.     
Once this is done, we restart this time keeping the corresponding bases
 at position $N_0$ from the open end at a distance $x$
 with the constraint that
the monomers at $N$ are at a distance $x'>N_0-x$,
to prevent rejoining in the already unzipped
$N_0$ monomers.
In this way we would sequence another $N_0$ monomers and so on.
We have verified this for models (Mi) and (Mii).   

In conclusion, we studied the ${\bf f}$ vs. ${\bf r}$  
characteristic curve  in the fixed stretch ensemble for simple models 
of DNA focusing on the base pairing energy only. 
We have seen that for a homo-DNA, the force difference between 
a native and the corresponding one mutation case when pulled to the same 
distance contains enough signature to locate the position of the 
mutation. This could be the basis of a differential force 
microscope to detect mutations.  For a heterogeneous DNA, the mutation point cannot
 be localized always as accurately as for homo-DNA.  Accuracy 
could be achieved by taking cognizance of the full features of the
$\delta f$ curve. 
We have shown that the differential force curve can be understood 
as due to the domain wall separating the zipped and the unzipped phases 
as the strands are pulled apart. Moreover, we found (Fig. 1) that the 
modulations in the force curve are connected to the local 
sequence. This holds promise of extension of our proposal to cases 
beyond point mutation.   
 
{\it Acknowledgements:} SMB thanks ICTP for hospitality. We thank 
A. Maritan and F. Seno for useful discussions.

\begin{figure}     
\begin{center}
\centerline{\psfig{figure=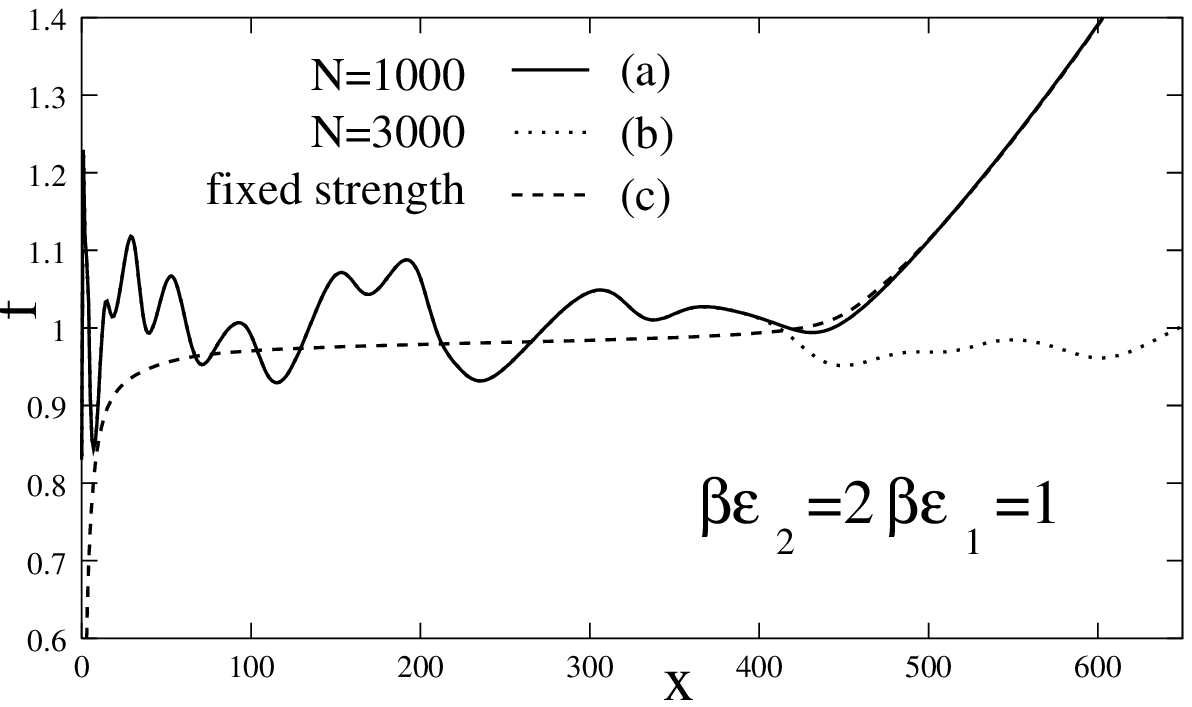,width=4.in}}
\end{center}
\caption {The force vs stretching distance curves for heterogeneous 
  DNAs (model (Mi)). The sequences are chosen 
randomly but both share the same sequence ($1000$ $\epsilon$'s) 
 from the open, pulled end. For the 
fixed stretch ensemble, curves (a) and (b), the pattern is  
identical  over a region of  $x$. Curve (c) is the fixed force 
 ensemble phase coexistence curve with finite-size effect.
The length of the unzipped part in units of base pairs is approximately
$x/y_0$ (see Eq. \ref{characteristic}).}                             
\label{fig:1} 
\end{figure}  
 
\begin{figure} 
\begin{center}
\centerline{\psfig{figure=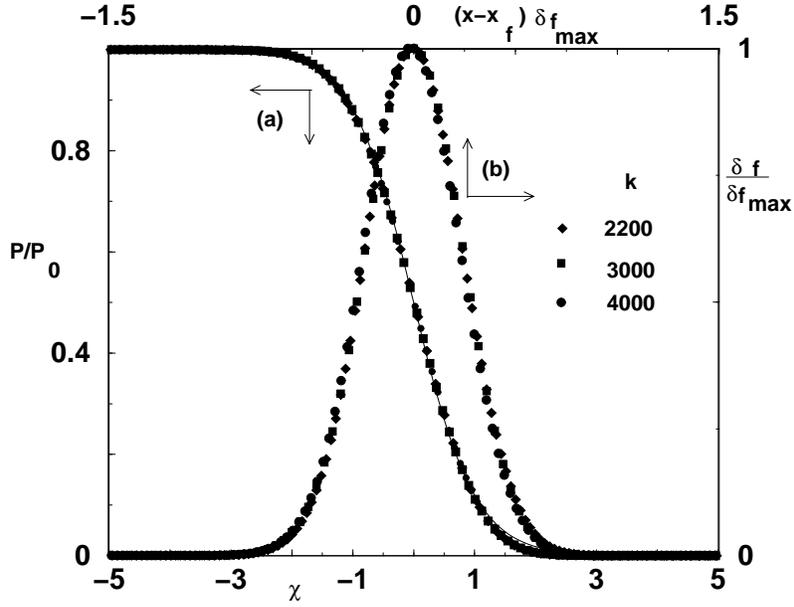,angle=270,width=4.1in}}
\end{center}
\caption{ (a) The collapse of $P(x,N|0,k)/P_0$ vs. 
$\chi$ where  
  $x_{\rm d}(k),w_{\rm d}(k)$ are obtained by fitting  
Eq. \ref{eq:3} (solid curve).   
 For clarity only three cases of $k$ are shown.  
 (b) The collapse plot of $\delta 
  f/\delta f_{\rm max}(k)$ vs $(x-x_f(k))\delta f_{\rm max}(k)$.  
 A similar collapse is found 
  even with  $x_{\rm d}(k)$ and $w_{\rm d}(k)$ of (a). 
 These are for model (Mii), homo-DNA, $N=5000$ , $d=1$
and $\beta\epsilon=2\beta\epsilon'=1.5$. Arrows point towards the
relevant axes.
}  
\label{fig:2} 
\end{figure}  

\newpage 
 
\begin{figure} 
\begin{center}
\centerline{\psfig{figure=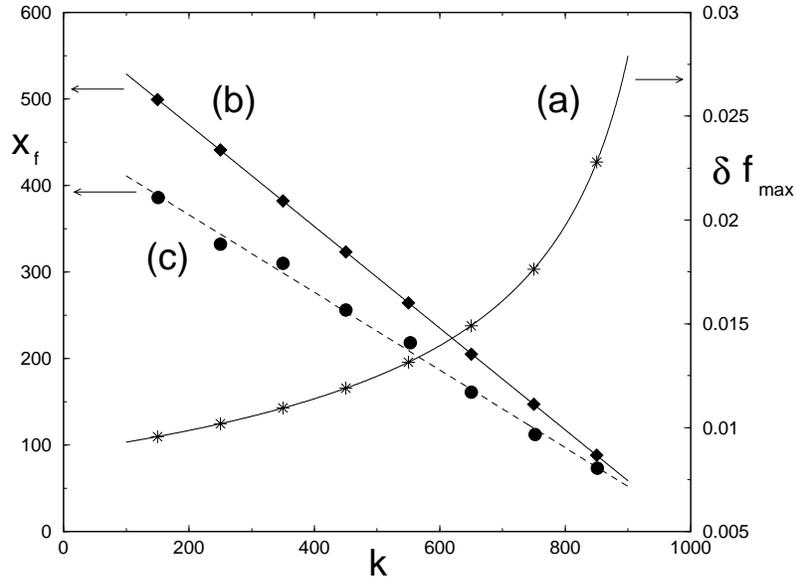,angle=270,width=4.1in}}
\end{center}
\caption{ The ``calibration'' curves $\delta 
  f_{\rm max}(k)$ and $x_f(k)$ for (a,b) a homogeneous DNA  and (c) 
  a heterogeneous DNA (only $x_f(k)$ is shown). The curves 
fitting the data according to Eq. \ref{results} are shown. 
Parameters are as in Fig. 1.  
} 
\label{fig:3} 
\end{figure} 

\begin{figure}   
\begin{center}
\centerline{\psfig{figure=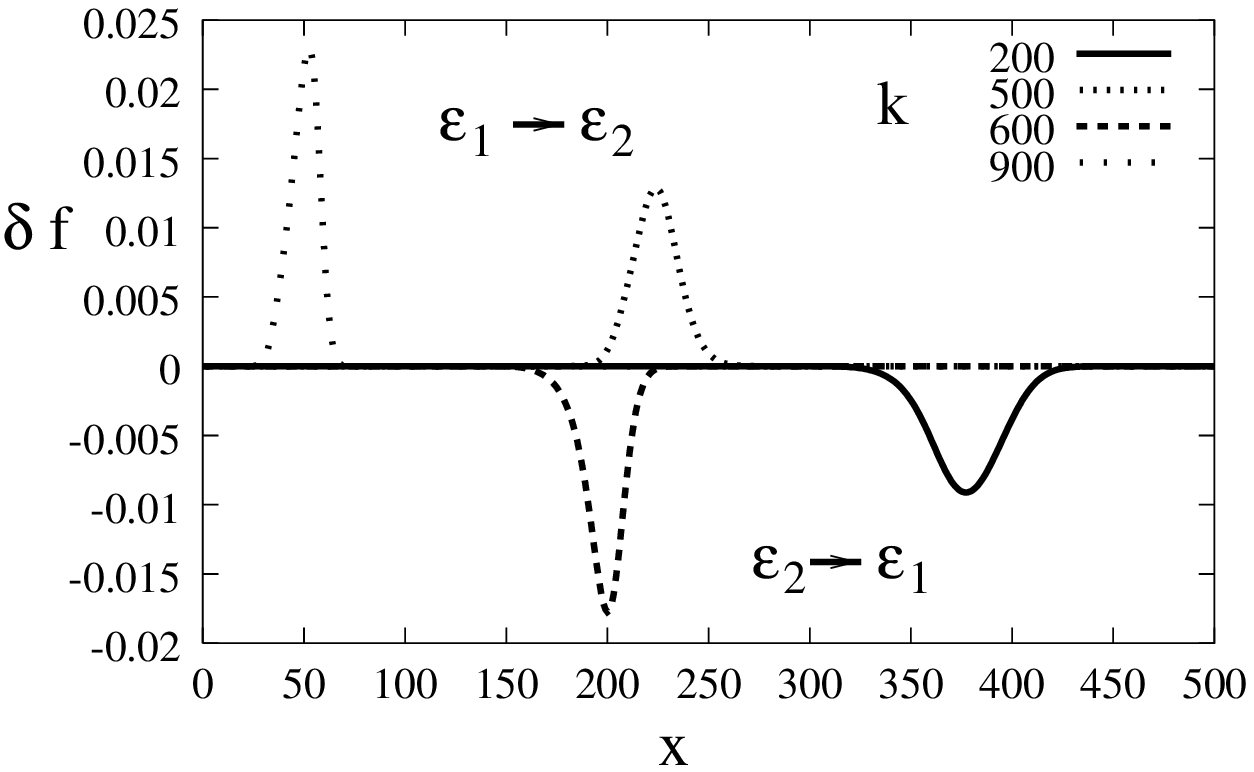,width=4.1in}}
\centerline{\psfig{figure=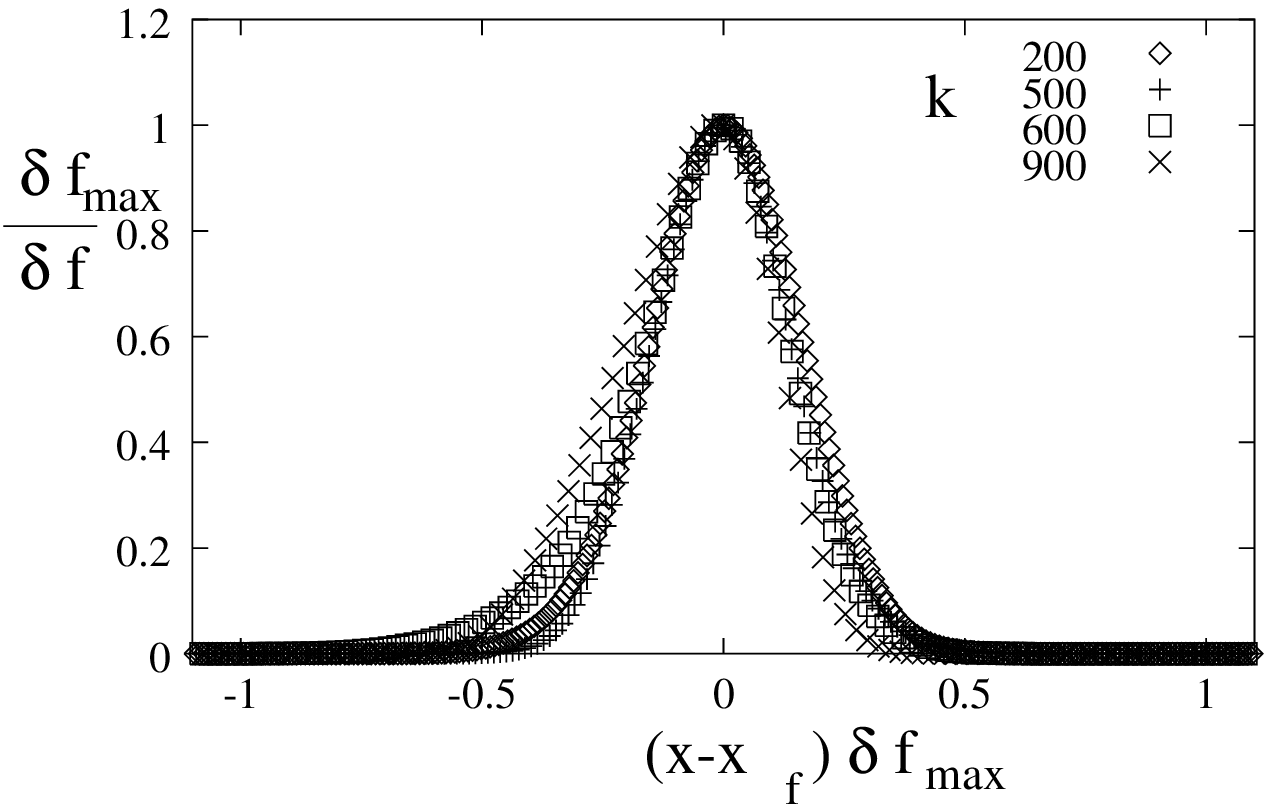,width=4.1in}}
\end{center} 
\caption{(a) The $\delta f$ vs $x$ curve for a heterogeneous DNA. The 
 sign of $\delta f$ gives the nature of the 
 mutation. (b) The collapse plot of the curves of (a) as in 
 Fig. \ref{fig:2}.The plots are for model (Mi), $N=1000$ 
and $\epsilon_2=2\epsilon_1=1$ (with $T=1$).  } 
\label{fig:4} 
\end{figure} 
 
\end{document}